\documentclass{aa}
\usepackage{graphicx}
\usepackage{txfonts}
\usepackage{textcomp}
\usepackage{natbib}
\bibpunct{(}{)}{;}{a}{}{,}

\begin{document}

\title{Mid-infrared observations of the transitional disks around DH\,Tau, DM\,Tau, and GM\,Aur\thanks{Based on observations made with ESO Telescopes at the La Silla or Paranal Observatories under programme ID 380.C-0326.}}
\author{Ch.~Gr\"afe\thanks{e-mail: cgraefe@astrophysik.uni-kiel.de} \inst{\ref{inst1}}\and S.~Wolf \inst{\ref{inst1}}\and V.~Roccatagliata \inst{\ref{inst2}}\and J.~Sauter \inst{\ref{inst1}}\and S.~Ertel \inst{\ref{inst1}} }
\institute{University of Kiel, Institute of Theoretical Physics and Astrophysics, Leibnizstrasse 15, 24098 Kiel, Germany \label{inst1}
  \and Space Telescope Science Institute, 3700 San Martin Drive, Baltimore, MD 21218, USA \label{inst2}}
\date{Received (date) / Accepted (date)}

\abstract
  {}
  {We present mid-infrared observations and photometry of the transitional disks around the young stellar objects \object{DH\,Tau}, \object{DM\,Tau}, and \object{GM\,Aur}, obtained with VISIR/VLT in N band. Our aim is to resolve the inner region and the large-scale structures of these transitional disks, carrying potential signatures of intermediate or later stages of disk evolution and ongoing planet formation.}
  {We use the simultaneously observed standard-stars as PSF reference to constrain the radial flux profiles of our target objects. Subtracting the obtained standard-star profile from the corresponding science object profile yields the flux residuals produced by the star-disk system. A detection threshold takes into account the background standard deviation and also the seeing variations during the observations to evaluate the significance of these flux residuals. On the basis of a simple model for the dust re-emission, we derive constraints on the inner radius of the dust disk.}
  {We spatially resolve the transitional disk around GM\,Aur and determine an inner-disk hole radius of $20.5^{+1.0}_{-0.5}$\,AU. The circumstellar disks around DH\,Tau and DM\,Tau are not spatially resolved but we are able to constrain the inner-disk hole radius to $<\!15.5^{+9.0}_{-2.0}$\,AU and $<\!15.5^{+0.5}_{-0.5}$\,AU, respectively. The performed photometry yields fluxes of $178\pm31$\,mJy for DH\,Tau, $56\pm6$\,mJy for DM\,Tau, and $229\pm14$\,mJy for GM\,Aur.}
  {}


\authorrunning{Ch.~Gr\"afe et al.}

\maketitle

\section{Introduction}

Classical T Tauri stars are low-mass, pre-main sequence stars surrounded by circumstellar disks. These disks are a natural outcome of the star formation process and are known to dissipate over time by means of stellar winds or photoevaporation by either the radiation of the central star or some external radiation \citep[e.g.,][]{2000prpl.conf..401H, 2001MNRAS.328..485C, 2007MNRAS.375..500A}, accretion onto the central star \citep{1998ApJ...495..385H}, grain growth and fragmentation \citep[e.g.,][]{2004Natur.432..479V, 2005A&A...434..971D}, and the formation of planets \citep[e.g.,][]{1986ApJ...309..846L, 1992ApJ...395L.115M, 1996Icar..124...62P, 1999ApJ...514..344B, 2002E&PSL.202..513B, 2002ApJ...568.1008C}. The way in which the disk material dissipates has important implications for the possibility of planet formation. The key to constraining the physical mechanisms responsible for disk evolution is the detection and detailed analysis of disks that are in the process of dispersing. The evolution of circumstellar disks is mostly studied in the infrared by analyzing stars that exhibit emission in excess of that expected from the stellar photosphere. In contrast to the case for high-resolution imaging in the optical/near-infrared, the contribution from the stellar photosphere is small compared to the disk emission at mid-infrared wavelengths.\par

Transitional disks are considered to represent an evolved stage of the accretion disks surrounding a Class~II source \citep{1984ApJ...287..610L, 1987IAUS..115....1L} and to be the precursors of the debris disks found around main-sequence stars \citep[e.g.,][]{2005ApJ...630L.185C, 2009RMxAC..35...33D}. The spectral energy distribution (SED) of transitional disks shows a deficit of emission in the near-infrared, steep slopes in the mid-infrared and an excess at far-infrared wavelengths that can be explained by inner-disk gap(s) or hole(s), while a prominent outer disk is still present \citep[e.g.,][]{2005ApJ...621..461D}. To date, the structure of the transitional disks has been mainly investigated by model fitting of their SED, which usually does not provide an unequivocal solution \citep{2005ApJ...621..461D, 2005ApJ...630L.185C, 2008ApJ...678L..59I, 2009RMxAC..35...33D}. Imaging of protoplanetary disks provides an independent check of SED modeling and a far less ambiguous method for studying the disk structure \citep[e.g.,][]{2003ApJ...591..267C, 2005A&A...437..525P, 2009ApJ...698L.169E, 2009ApJ...698..131H, 2009A&A...505.1167S}.\par

Investigating the evolution and thus dissipation of circumstellar disks is crucial for our understanding of planetary system formation. Various efforts to spatially resolve the inner structure of transitional disks have been undertaken \citep[e.g. with the Submillimeter Array by][]{2009ApJ...704..496B}. However, owing to their limited sensitivity, submillimeter and millimeter observations only provide information about the large reservoirs of cold dust in the outer regions so far. Thus, high-resolution mid-infrared imaging of transitional disks is essential to obtain information about the warm dust in the innermost regions of the disk, to detect inner-disk clearings as signatures of the disk dissipation mechanisms and to investigate the structure close to the star and therefore the regions where planets potentially form \citep[e.g.,][]{2007P&SS...55..569W}.\par
\begin{table*}
\caption{Observing log}
\label{tab1}
\centering
\begin{tabular}{l c c c c c}
\hline\hline
Target		& RA (J2000)	& Dec (J2000)	& Observation Date	& Observation Time	& Exposure Time\\
		& [hh mm ss]	&  [dd mm ss]	&			& [hh:mm]		& [s]\\
\hline
DH\,Tau		& 04 29 41.56	& $+$26 32 58.3	& 2007 Oct 24		& 07:27			& 883.2\\
DM\,Tau		& 04 33 48.72	& $+$18 10 10.0	& 2007 Oct 20		& 06:26 / 07:30		& 2$\times$1766.4\\
GM\,Aur		& 04 55 10.98	& $+$30 21 59.5	& 2007 Oct 20		& 08:30			& 353.3\\
\hline
\end{tabular}
\end{table*}

\begin{table*}
\caption{Properties of the observed standard-stars}
\label{tab2}
\centering
\begin{tabular}{l c c c c}
\hline\hline
Target		& RA (J2000) [hh mm ss]	& Dec (J2000) [dd mm ss]	& Spec. Type	&$\rm m_{SIV-Filter}$ [mag]\\
\hline
HD\,31421	& 04 56 22.27		& $+$13 30 52.1			& K2III		& 11.38\\
HD\,27482	& 04 21 15.26		& $+$27 21 00.9			& K5III		& 6.12\\
\hline
\end{tabular}
\tablefoot{From flux catalog of mid-infrared standard-stars for VISIR imager filters based on \citet{1999AJ....117.1864C}.}
\end{table*}
In the mid-infrared, various efforts to spatially resolve circumstellar disks have been undertaken. For example, \citet{2008ApJ...689..539C} and \citet{2011MNRAS.410....2C} presented resolved mid-infrared images of the disk of HD\,191089 and HD\,181327, respectively. The circumstellar disk of the well-studied Herbig Ae/Be star AB\,Aur was investigated by \citet{2003ApJ...591..267C} and \citet{2005A&A...437..525P} in the mid-infrared.\par

In contrast to previous mid-infrared imaging of circumstellar disks, our observations and data analysis presented here reach the limit of angular resolution of today's single-dish, ground-based telescopes. In the case of AB\,Aur, which emits a N band flux of $>\!20$\,Jy, structures with a radius of $\sim\!\!280$\,AU were detected \citep{2003ApJ...591..267C, 2005A&A...437..525P}. We investigate sources that are $\sim\!\!100$ times fainter and structures that have more than ten times smaller radii. Although we are using one of the best observing facilities available, the properties of the observed objects necessitate a new data analysis procedure, which is described in Sect. 4. An even higher resolution in this wavelength range can only be achieved by means of interferometric observations, e.g., with MIDI/VLTI \citep[e.g.][]{2009A&A...502..367S}. However, such observations are still limited to the mid-infrared brightest circumstellar disks.\par

\section{Target objects}

Our target objects are three transitional disks that were selected based on their 10\,\textmu m flux from a list of most well-studied potential transitional disks \citep{2004ApJS..154..443F, 2006ApJ...648..484H, 2006ApJS..165..568F}. These three objects are DH\,Tau, DM\,Tau, and GM\,Aur, which are all located in the Taurus/Auriga star-forming region at a distance of $\sim\!\!140$\,pc \citep{1994AJ....108.1872K}. Their SEDs \citep[see Fig.~1 in][]{2006ApJS..165..568F} show significant infrared excess emission, which is at least ten times higher than the stellar emission in the mid-infrared. As it is typical of transitional disks, their SEDs also have a flux deficiency in the near-infrared with increasing slopes in the mid-infrared wavelength range, indicating an inner radius of the dust disk much larger than the dust sublimation radius.\par

In the following, our target stars are briefly characterized:

\begin{enumerate}
  \item DH\,Tau is a Class II source of spectral type M1 \citep{2009ApJS..180...84W}. \citet{2005ApJ...620..984I} detected a young brown dwarf companion to the transitional disk around DH\,Tau, separated by $2.3\arcsec$, corresponding to a spatial distance of 330\,AU if the orbit is seen face-on.

  \item DM\,Tau is a low-mass, pre-main sequence star that does not show any significant excess emission below 8\,\textmu m, but is surrounded by a still accreting transitional disk with a mass accretion rate slightly lower than the average of disks in Taurus ($\dot{M}_{\rm average}=10^{-7}-10^{-8}\,M_{\sun}$/yr) \citep{1998ApJ...495..385H, 2001ApJ...556..265W, 2009ApJ...699..330S}. Its inner-disk hole radius is expected to be 3\,AU with little dust inside \citep{2005ApJ...630L.185C, 2010ApJ...710..265P}.

  \item GM\,Aur is a 3\,Myr old K7 T\,Tauri star that is surrounded by a transitional disk and is also known to be still in its accreting phase \citep{2005ApJ...630L.185C, 2008A&A...490L..15D}. Its SED indicates the presence of an inner-disk hole of radius $20-24$\,AU that is partially filled with optically thin dust \citep{2005ApJ...630L.185C, 2009ApJ...698..131H, 2010ApJ...710..265P}. \citet{2008A&A...490L..15D} and \citet{2009ApJ...698..131H} suggest that the inner-disk hole plausibly results from the dynamical influence of a planet on the disk material.
\end{enumerate}

\section{Observations and data reduction}

\begin{table}[b]
\caption{Observation properties of the observed standard-stars}
\label{tab3}
\centering
\begin{tabular}{l c c c c}
\hline\hline
Target		& Obs. Date	& Obs. Time	& Exposure	& N Band	\\
		& [2007]	& [hh:mm]	& Time [s]	& Seeing [$\arcsec$] \\
\hline
HD\,31421	& Oct 20	& 06:16		& $61.44$	& $0.29$	\\
HD\,31421	& Oct 20	& 07:22		& $61.44$	& $0.26$	\\
HD\,27482	& Oct 20	& 08:20		& $61.44$	& $0.29$	\\
HD\,27482	& Oct 24	& 07:16		& $61.44$	& $0.29$	\\
\hline
\end{tabular}
\end{table}

The mid-infrared observations of our three targets were carried out on 2007 October 20 and 24, as described in Table~\ref{tab1}, using VISIR \citep{2004Msngr.117...12L} in the SIV filter ($\lambda_c=$\,10.49\,\textmu m, $\Delta\lambda=$\,0.16\,\textmu m). VISIR is the ESO/VLT mid-infrared imager and spectrograph, composed of an imager and a long-slit spectrometer covering several filters in N and Q bands. To achieve the highest possible angular resolution, the small field objective was used (0.075\arcsec/pixel). A standard chop/nod-procedure to remove sky and telescope emission with a chop of 8$\arcsec$ in the north-south direction and a perpendicular 8$\arcsec$ nod were applied for the observations. Calibration observations of standard-stars within a few degrees of the science object were taken immediately before the science observations. The standard-stars were chosen from the mid-infrared spectrophotometric standard-star catalog of the VLT based on \citet{1999AJ....117.1864C} (see Table~\ref{tab2}). These standard-stars were used to characterize the point spread function (PSF) for comparison with the science sources to detect any signs of a circumstellar disk. All observations were executed in service mode. Raw data were reduced using the VISIR data reduction pipeline provided by ESO. A full-resolution background map was created for every observed object using the Source Extractor software \citep{1996A&AS..117..393B}. This map was subtracted from the image to remove the background. This approach provides a more reliable estimate of the background than the assumption of a constant background.\par

\section{Data analysis}

\subsection{Radial flux profile}

The key components of each of our science targets are the central star and the surrounding disk. In contrast to the brightness profile of the corresponding reference (standard-) star, the profiles of the science targets are extended by the circumstellar disk. The standard-stars, which are considered to be point sources, are used as PSF references (see Tables~\ref{tab2} and \ref{tab3}). To obtain the flux profile of a science object and a corresponding standard-star, we carry out a radial averaging of each object. For this purpose, the images are resized by a factor of ten using bilinear interpolation. The center of each object is determined by fitting a two-dimensional Gaussian profile. The radial averaging is then performed with respect to the obtained object center and yields as a result the radial flux profile of the object. Assuming radial symmetry, this averaging method yields a much better flux profile than only one slice through the object and also considerably improves the signal-to-noise ratio (hereafter S/N) by a factor of $\sqrt{N}$, where $N$ is the number of pixels in every annulus. Radially symmetric disk images can of course only be expected if the disks are seen in face-on orientation. The effect of a disk inclination other than face-on on our results is discussed in Sect. 4.3.\par

Owing to the small chopping and nodding throw in the observations, the observed source is always present in the field of view of the instrument such that the source is detected four times in the final reduced image (as two positive and two negative images). To increase the S/N by a factor of two, the radial averaging is accomplished for all four observations of the object in the final image and averaged afterwards. This profile is then normalized to a peak flux of 1. Subtracting the standard-star profile from the corresponding science object profile yields the flux residuals produced by the extended star-disk system.\par

\subsection{Detection threshold}

At the next step, one has to investigate whether the derived flux residuals are significant. Therefore, we define a detection threshold such that any residual flux above this threshold is caused primarily by an extension of the star-disk system, i.e., a spatially resolved circumstellar disk surrounding the star.\par

Our detection threshold takes into account both the standard deviation in the background flux ($\sigma_{\rm bg}$) of the science image and the seeing variations during the observations of the standard-star and science object. The background noise is calculated from the full-resolution background noise map created using Source Extractor. We impose the constraint that the residual flux must be higher than $3\sigma_{\rm bg}$. Owing to the radial averaging, the background noise of the flux profile depends on the radius and declines with increasing radius.\par

Standard-stars are always observed before and/or after the science object to have nearly the same observing conditions, thus for point sources nearly the same full-width-at-half-maximum (FWHM). However, the FWHM depends on the seeing, which varies during observations. If not properly taken into account, variations in seeing can cause a difference signal similar to the case of an extended star-disk system, thus mimic a spatially resolved disk. For our data analysis, it is therefore crucial to determine the seeing variations during the observations of the standard-star and science object. This is difficult because information about the seeing during observations of the science object at the observed wavelength is unavailable. Only the FWHM of the standard-star is known from which the seeing at both the observation time and the observed wavelength must then be determined (see Table~\ref{tab3}). To estimate the seeing variations, we analyze observations of our standard-star HD\,31421, which was repeatedly observed over several hours before and after the observations of a scientific target. Therefore, we use the observation with the highest and the smallest FWHM of this standard-star, i.e., at the worst and the best seeing conditions and determine the radial flux profile of both as described above. Assuming that the seeing has a normal distribution during the observations, the seeing variations can now be calculated as the half difference of these two profiles. This method yields a radial profile reflecting a conservative estimate of the seeing variations during the observations. The derived seeing variations are adopted for all observed science objects.\par

\begin{figure}[b]
 \centering
  \includegraphics[width=0.50\textwidth]{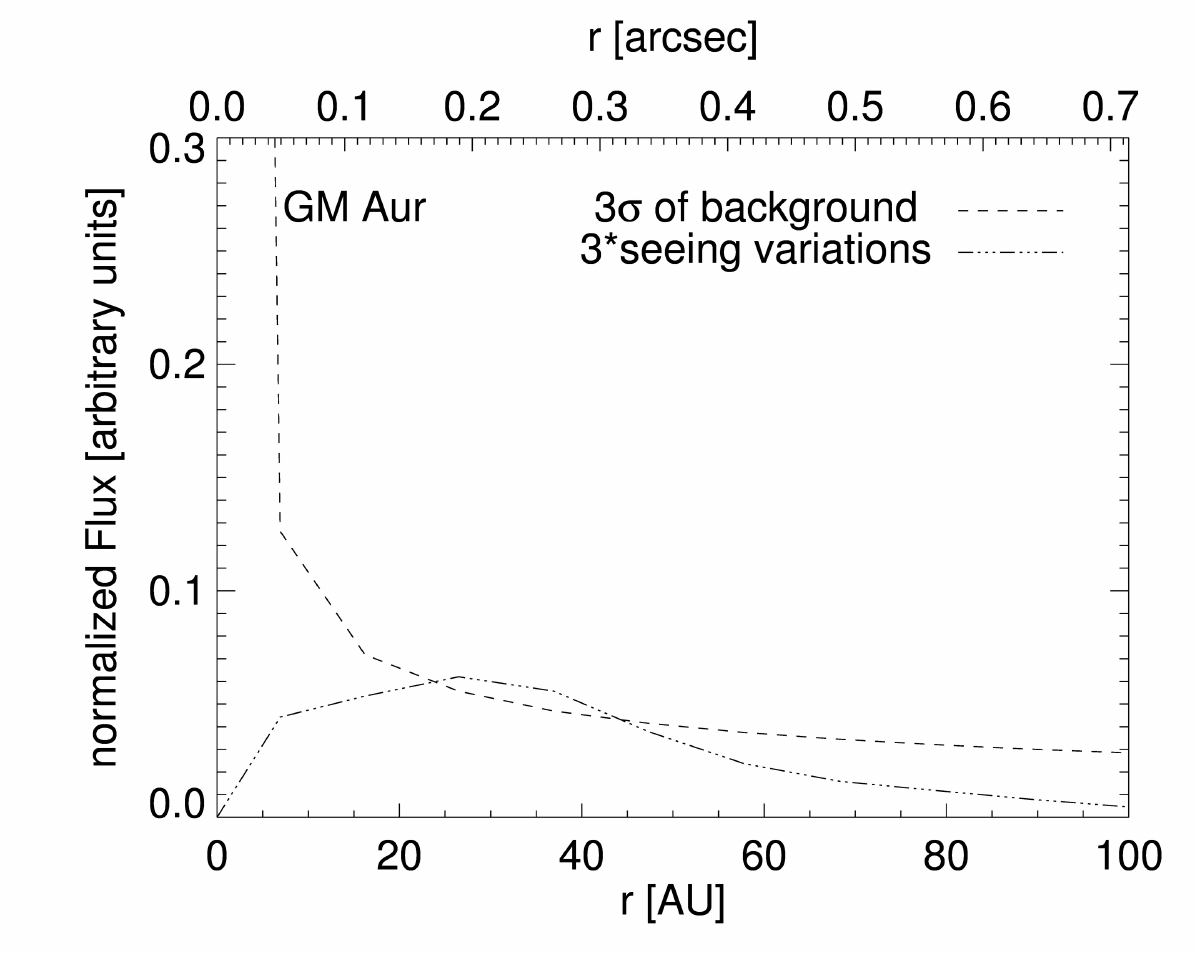}
\caption{The contribution of the two factors to the detection threshold.}
\label{Fig4}
\end{figure}

To increase the statistical significance of the derived flux residuals, we also insist that they be three times larger than the derived seeing variations. Our detection threshold is now defined as the sum of $3\sigma_{\rm bg}$ and three times the seeing variations. The contribution of the two components to this threshold is exemplarily represented for GM\,Aur in Fig.~\ref{Fig4}, where it is clearly visible that the seeing variations are a significant part of the threshold, hence must be reliably determined.\par

\subsection{Model}

We present a simple model to determine the inner-disk hole radius of our objects, provided that they are spatially resolved in our observations. However, our approach also allows us to estimate an upper limit to the inner-disk hole radius of our transitional disks that could not be spatially resolved.\par

Our model consists of a star surrounded by a disk observed at a certain inclination. To allow comparison with the observations, the image of this model was convolved with the measured PSF. This PSF is the arithmetic mean of the four observations of the observed standard-star in the final image, which are caused by the chop/nod technique. For every model of the science objects, the corresponding standard-star is used.\par

The inner disk rim, which dominates the mid-infrared surface brightness distribution of the disk, is represented by an inclined circle with a radius equal to the gap/hole radius of the transitional disk whose inclination equals the known inclination of the observed disk. The star is located in the center with a star/disk flux ratio obtained from the observed SED \citep{2006ApJS..165..568F, 2009ApJ...698..131H} (see Table~\ref{tab4}). As described in Sect. 4.1, the flux profiles of the PSF-convolved object and the measured PSF are obtained using a radial averaging method. They are normalized to a peak flux of 1. The radial profile of the measured PSF is then subtracted from the convolution profile to obtain the flux residuals arising from the extended star-disk system. These flux residuals and the determined detection threshold yield the model to the corresponding observation.\par

\begin{table}[h]
\caption{Star/disk flux ratio at 10\,\textmu m of the observed objects. For details, see Sect. 4.3.}
\label{tab4}
\centering
\begin{tabular}{l c}
\hline\hline
Target		& Star/disk flux ratio	\\
\hline
DH\,Tau		& 0.111$\pm$0.018	\\
DM\,Tau		& 0.125$\pm$0.015	\\
GM\,Aur		& 0.081$\pm$0.013	\\
\hline
\end{tabular}
\end{table}

\subsection{Photometry}

The N band fluxes of the three observed objects are determined using the VISIR photometry package by \citet{2010A&A...515A..23H}. For photometry, the science targets can be considered as point sources such that each science and calibrator beam is fitted by a two-dimensional Gaussian. The fluxes are then obtained by calculating conversion factors from the integrated intensity of the Gaussian fits of the calibrators and using these factors to evaluate the respective integrated Gaussian intensity of the science targets.\par

\section{Results and discussion}

The resulting constraints on the inner-disk hole radii and the photometry of the observed transitional disks are compiled in Table~\ref{tab5}. In the following, we discuss the analysis of the individual objects in detail.\par

\begin{table}[ht]
\caption{Constraints on the inner-disk hole radii and photometry of the observed transitional disks}
\label{tab5}
\centering
\begin{tabular}{l c c}
\hline\hline
Target		& Inner-disk hole radius [AU]	&	Flux [mJy]\\
\hline
DH\,Tau		& $<\!15.5^{+9.0}_{-2.0}$	&	$178\pm31$\\
DM\,Tau		& $<\!15.5^{+0.5}_{-0.5}$	&	$56\pm6$\\
GM\,Aur		& $20.5^{+1.0}_{-0.5}$		&	$229\pm14$\\
\hline
\end{tabular}
\tablefoot{The uncertainties in the inner-disk hole radii are caused mainly by the uncertainty in the disk inclination (see Sect. 5).}
\end{table}

\begin{figure*}[t]
 \centering
  \includegraphics[width=0.50\textwidth]{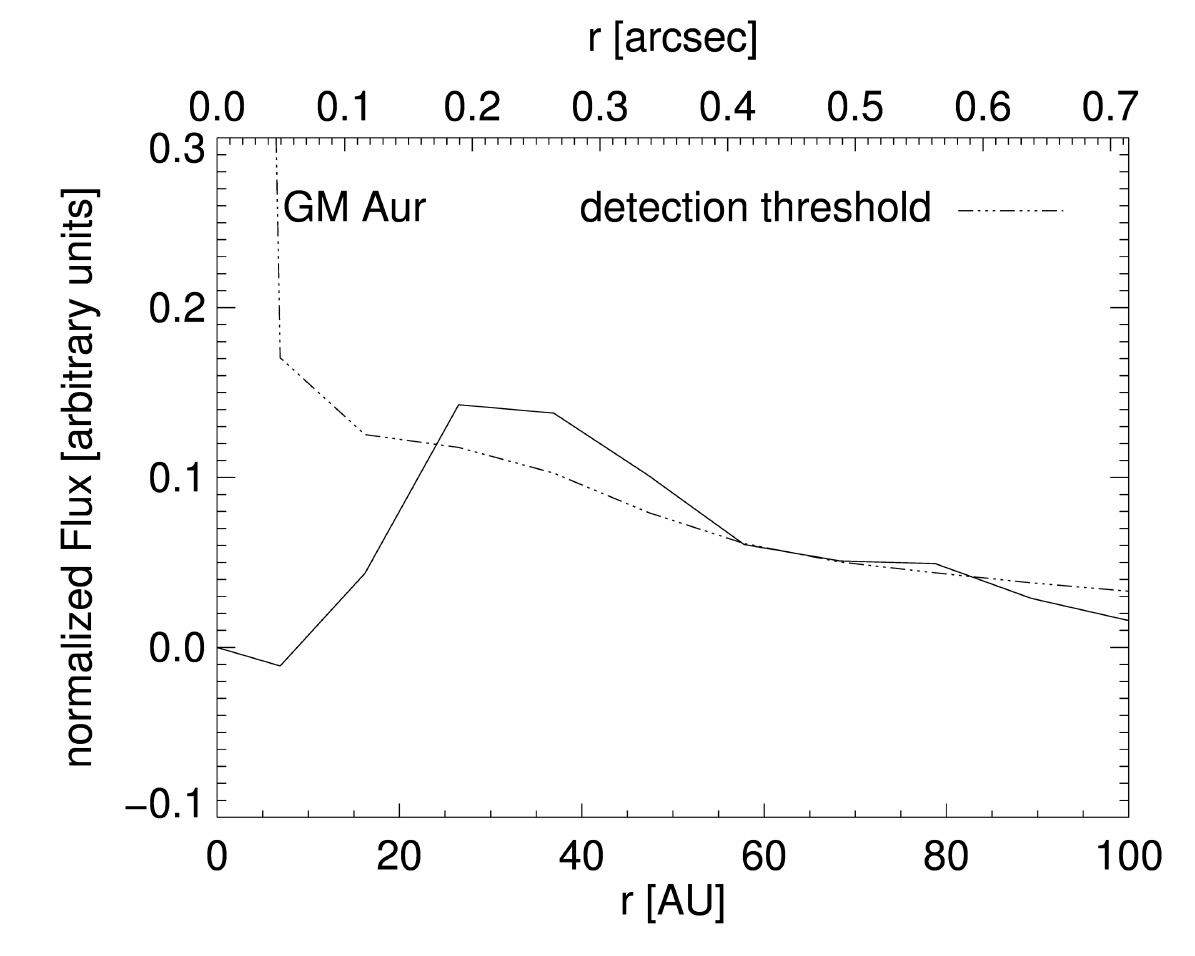}
  \hfill
  \includegraphics[width=0.49\textwidth]{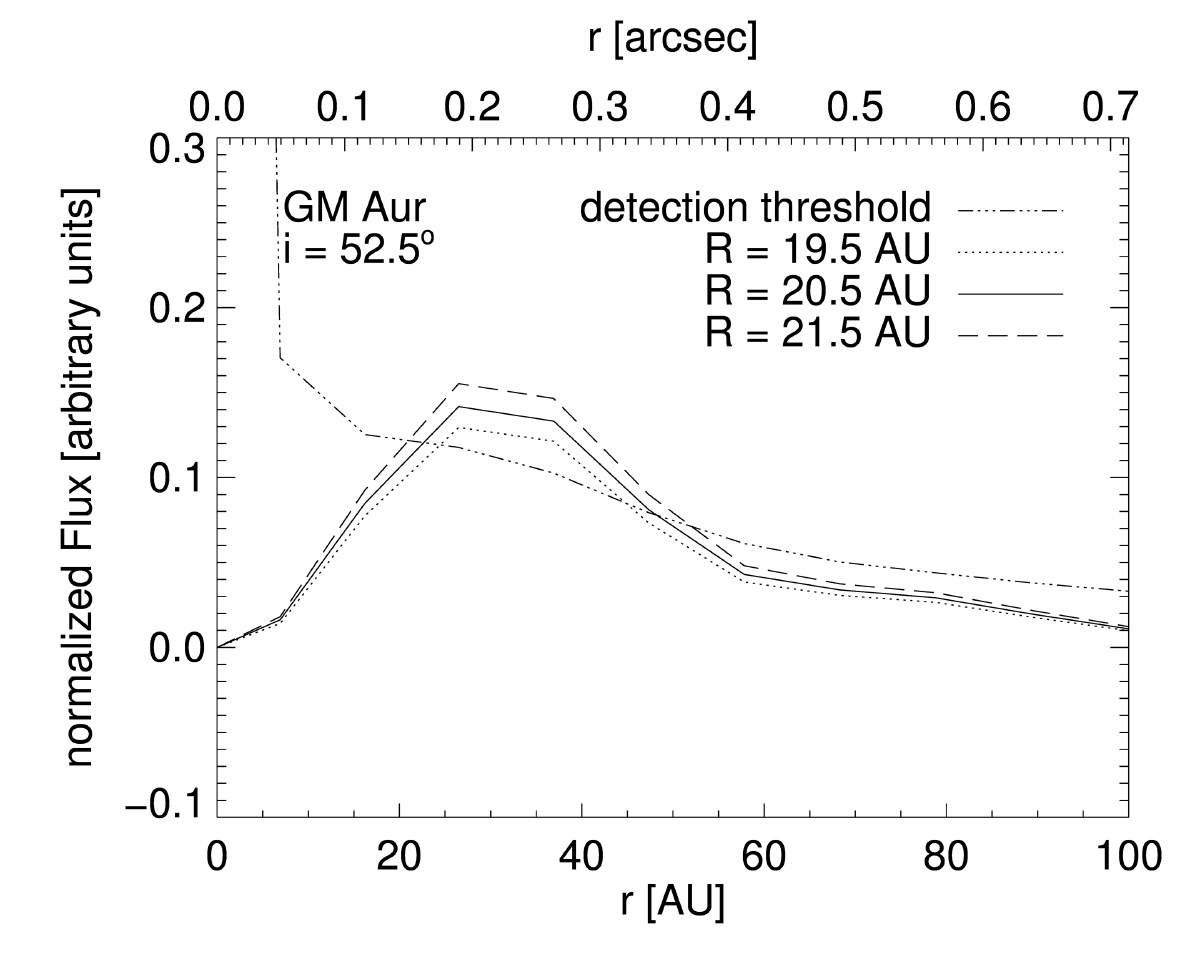}
\caption{\emph{Left}: Observed flux residuals of GM\,Aur including the determined detection threshold. \emph{Right}: Modeled flux residuals for several inner-disk hole radii including the detection threshold.}
\label{Fig1}
\end{figure*}

\begin{figure*}[t]
 \centering
  \includegraphics[width=0.50\textwidth]{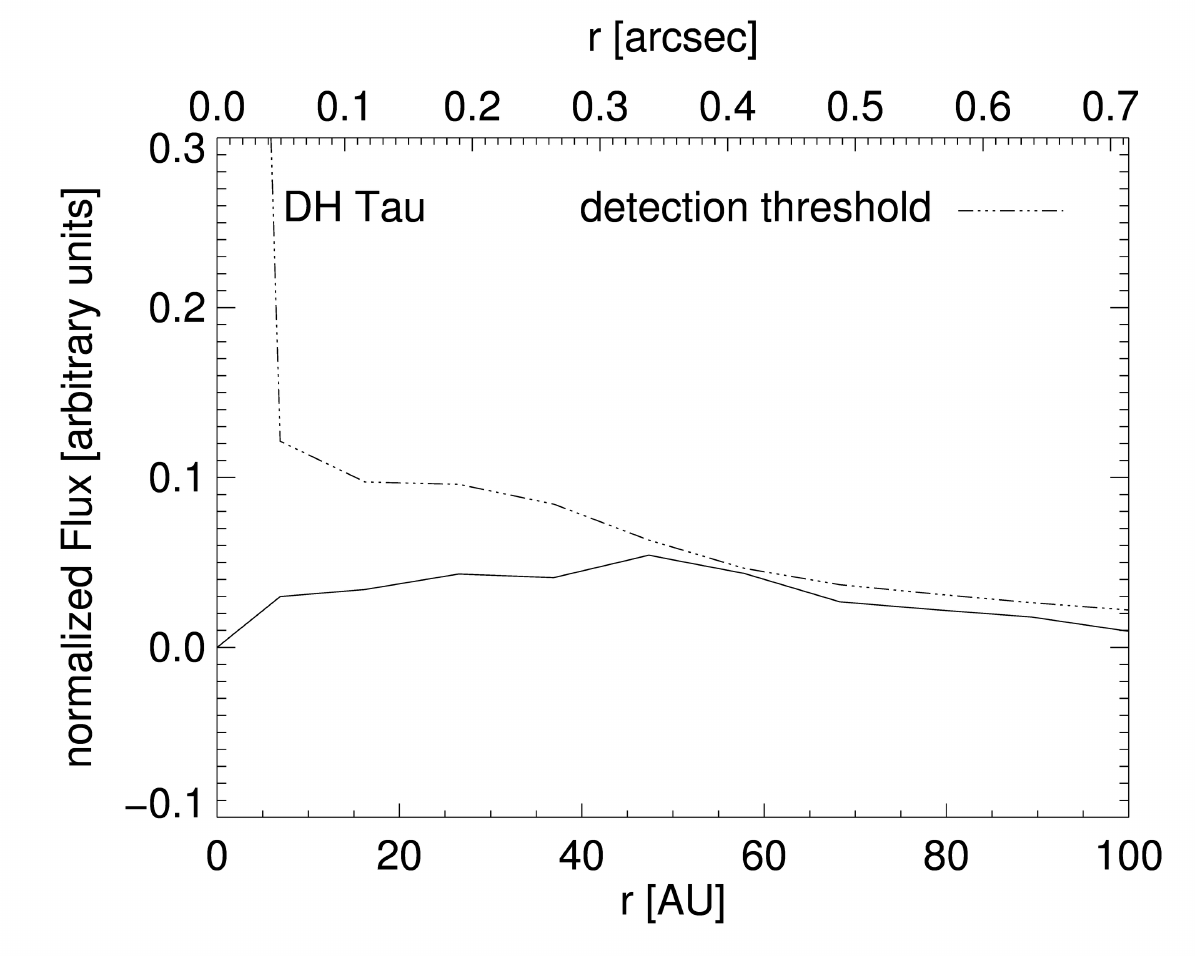}
  \hfill
  \includegraphics[width=0.49\textwidth]{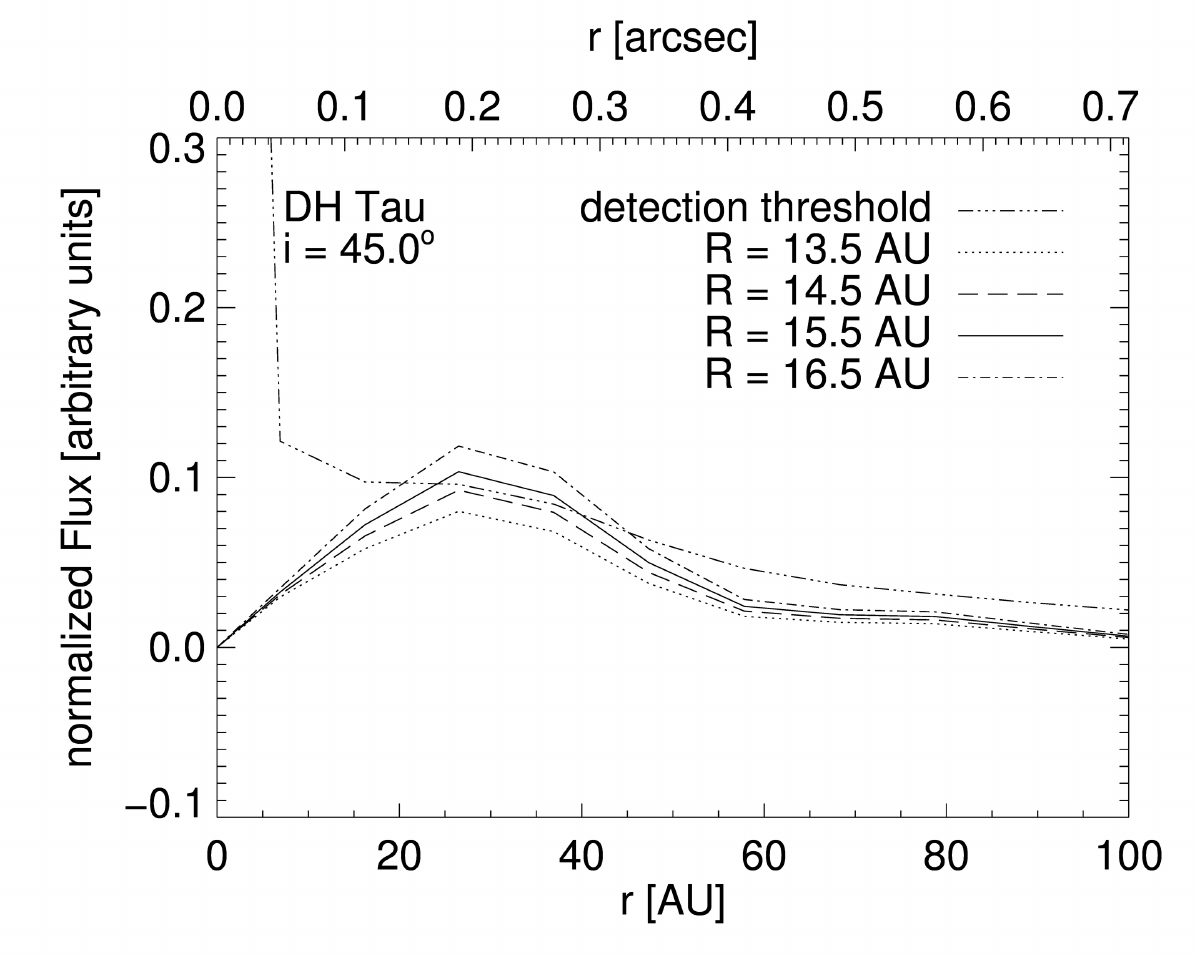}
\caption{\emph{Left}: Observed flux residuals of DH\,Tau including the determined detection threshold. \emph{Right}: Modeled flux residuals for several inner-disk hole radii including the detection threshold.}
\label{Fig2}
\end{figure*}

\begin{figure*}[t]
 \centering
  \includegraphics[width=0.50\textwidth]{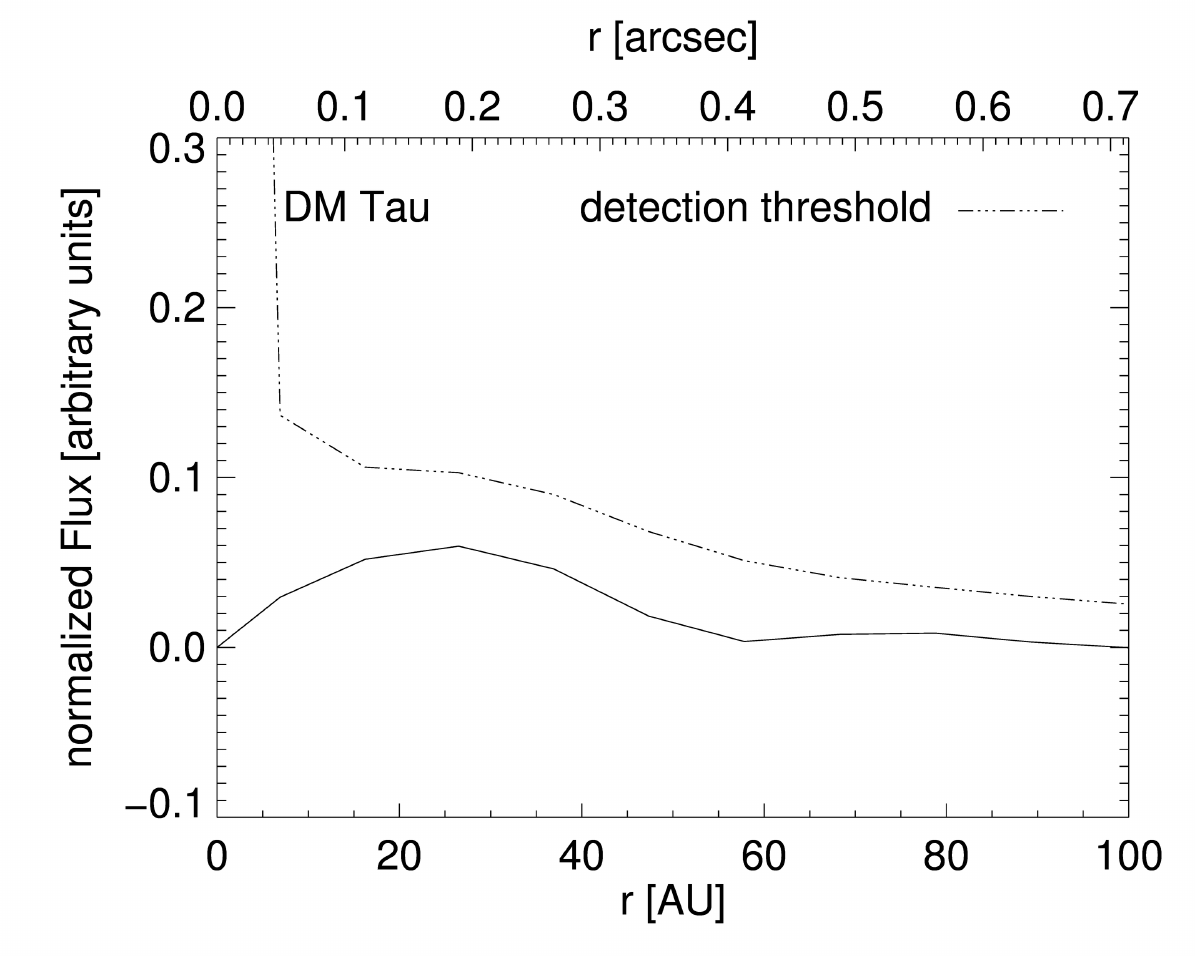}
  \hfill
  \includegraphics[width=0.49\textwidth]{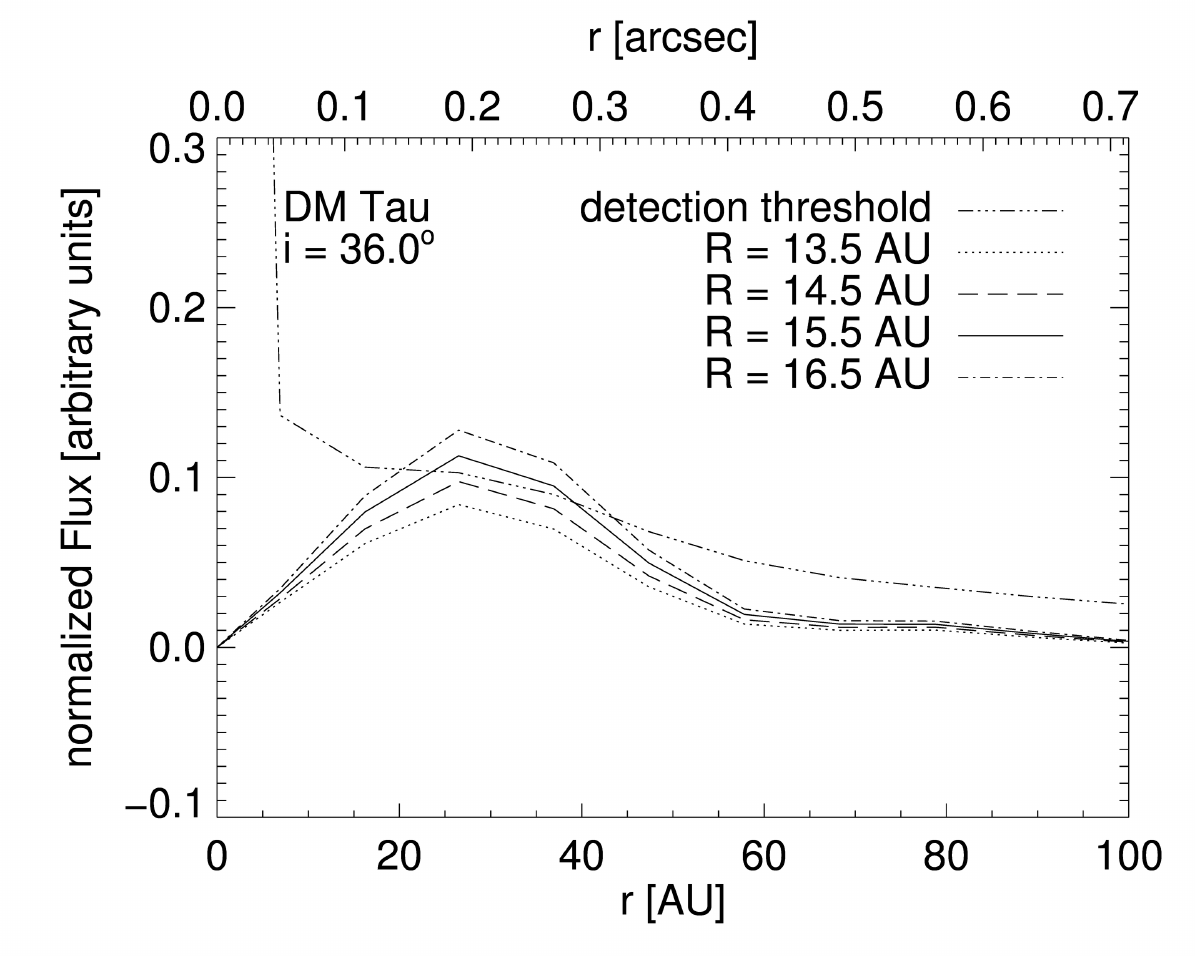}
\caption{\emph{Left}: Observed flux residuals of DM\,Tau including the determined detection threshold. \emph{Right}: Modeled flux residuals for several inner-disk hole radii including the detection threshold.}
\label{Fig3}
\end{figure*}

For GM\,Aur, we are able to detect the extension of the star-disk system caused by its circumstellar disk (see Fig.~\ref{Fig1}, left). Estimates of the disk inclination of GM\,Aur range from 49$\degr$ to 56$\degr$ \citep{2000ApJ...545.1034S, 2005ApJ...630L.185C, 2008A&A...490L..15D, 2009ApJ...698..131H, 2009ApJ...701..260I}. Comparing the observation with our model, we can constrain the inner-disk hole radius of GM\,Aur's transitional disk to $20.5^{+1.0}_{-0.5}$\,AU (see Fig.~\ref{Fig1}, right), where the smaller value represents the radius for the smaller inclination and the larger value the radius for the larger inclination. This size is comparable to the observational seeing in the N band as a disk hole of radius $20.5$\,AU at $140$\,pc amounts to $\sim\!0.29\arcsec$ across. The inner-disk hole radius derived from our observations is in very good agreement with the values of $\sim\!\!20$\,AU and $\sim\!\!24$\,AU found by \citet{2009ApJ...698..131H} and \citet{2005ApJ...630L.185C}, respectively.\par

In Figs.~\ref{Fig2} and \ref{Fig3}, we present the results and models for DH\,Tau and DM\,Tau. For both objects, the difference in the object and standard-star flux profiles is below the detection threshold, i.e., no significant extension of the star-disk system is detected. Using our model, we can investigate at which inner-disk hole radius we would have detected a significant extension of the star-disk system, hence measure an upper limit to the inner-disk hole radius. The disk inclination of DM\,Tau ranges from 32$\degr$ to 40$\degr$ \citep{2000ApJ...545.1034S, 2005ApJ...630L.185C, 2010ApJ...710..265P}. We therefore infer that DM\,Tau's inner-disk hole radius is smaller than $15.5^{+0.5}_{-0.5}$\,AU, which thus agrees with \citet{2005ApJ...630L.185C} who expected an inner-disk hole radius of $3$\,AU. For DH\,Tau, no estimate of the disk inclination has yet been made. Hence, using our model we determine an upper limit to the disk hole radius of $15.5^{+9.0}_{-2.0}$\,AU, where the lower limit corresponds to when the disk is seen face-on and the upper limit to an edge-on orientation. However, a close to edge-on orientation of this disk can be excluded because the mid-infrared bright inner-disk region would not then be visible through the optically thick disk. No complementary constraints on the inner-disk hole radius are yet available for DH\,Tau.\par

When considering deviations in both the radial flux profiles and the detection threshold, besides the detection threshold per se, the error bars are far too small to be visible in the plots. The error in the radial profiles is the background noise, which is implemented in the detection threshold. The error in the second part of this threshold, i.e. in the seeing variations, is half times the sum of the quantity $\sigma_{\rm bg}$ of the standard-star for the worst seeing conditions plus the one for the best. Owing to their high S/N and the radial averaging, the background noise of the standard-stars is negligibly small. Therefore, this also applies to the error in the seeing variations. The contribution of $3\sigma_{\rm bg}$ and three times the seeing variations to the detection threshold can be seen in Fig.~\ref{Fig4}.\par

The performed N band photometry of our three observed transitional disk objects yields fluxes of $178\pm31$\,mJy for DH\,Tau, $56\pm6$\,mJy for DM\,Tau, and $229\pm14$\,mJy for GM\,Aur. They are the first imaging photometric measurements of these objects at this wavelength. As a complement to already known fluxes at other wavelengths, they will be useful to constrain the SEDs of these three objects and model their circumstellar disks.\par

Spatially resolved 2\,\textmu m Keck Interferometer observations of DM\,Tau and GM\,Aur presented by \citet{2010ApJ...710..265P} indicate that there is a significant K band flux contribution from extended hot circumstellar matter very close to the star. This is consistent with previous SED modeling of DM\,Tau and GM\,Aur \citep{2005ApJ...630L.185C, 2009ApJ...698..131H} that also inferred that the amount of dust in the inner few AU is too small to contribute significantly to the mid-infrared flux. Nevertheless, our results help to reduce the ambiguities of the SED modeling and allow us to conclude that this hot dust has to be located inside our measured inner-disk hole radii. Higher-resolution observations, which will potentially be feasible with the Atacama Large Millimeter Array, will allow us to spatially resolve substructures within $\sim\!\!10$\,AU and thus investigate whether the inner thin disk traced in K band is connected to the ``outer'' disk or a gap exists.\par

\begin{acknowledgements}
This work is supported by the DFG through the research group 759 ''The Formation of Planets: The Critical First Growth Phase''. We are grateful to the anonymous referee for providing useful suggestions that greatly improved this paper.
\end{acknowledgements}

\bibliographystyle{aa}
\bibliography{bibfile}

\end{document}